# A TAXONOMY OF PERFORMANCE ASSURANCE METHODOLOGIES AND ITS APPLICATION IN HIGH PERFORMANCE COMPUTER ARCHITECTURES

Hemant Rotithor

Microprocessor Architecture Group (IDGa)
Intel Corporation Hillsboro, OR, 97124, USA
hemant.g.rotithor@intel.com

## ABSTRACT

*This paper presents a systematic approach to the complex problem of high confidence performance assurance of high performance architectures based on methods used over several generations of industrial microprocessors. A taxonomy is presented for performance assurance through three key stages of a product life cycle-high level performance, RTL performance, and silicon performance. The proposed taxonomy includes two components-independent performance assurance space for each stage and a correlation performance assurance space between stages. It provides a detailed insight into the performance assurance space in terms of coverage provided taking into account capabilities and limitations of tools and methodologies used at each stage. An application of the taxonomy to cases described in the literature and to high performance Intel architectures is shown. The proposed work should be of interest to manufacturers of high performance microprocessor/chipset architectures and has not been discussed in the literature.*

## 1. INTRODUCTION

Phases in the design of a high performance architecture include: generating ideas for performance improvement, evaluation of those ideas, designing a micro-architecture to implement ideas, and building silicon implementing key ideas. At each stage, potential performance improvements need to be tested with high confidence. Three stages of developing a high performance architecture correspond to three levels of abstraction for performance assurance- *high level (HL) performance, RTL performance, and silicon performance.* Performance assurance consists of performance analysis of key ideas at a high level, performance correlation of the implementation of a micro-architecture of these ideas to high level analysis/expectations, and performance measurement on the silicon implementing the micro-architecture. Examples of high performance architectures include-microprocessors, special purpose processors, memory controller and IO controller chipsets, accelerators etc.

A successful high performance architecture seeks major performance improvement over previous generation and competitive products in the same era. Significant resources are applied in developing methodologies that provide high confidence in meeting performance targets. A high performance architecture may result in several products with different configurations, each of which has a separate performance target. For example, a CPU core may be used in server, desktop, mobile products with different cache sizes, core/uncore frequencies, number of memory channels/size/speed. A performance assurance scheme should provide high confidence in performance of each product. We propose a generalized taxonomy of performance assurance methods that has been successfully deployed for delivering high performance architectures over several generation of CPUs/chipsets. The proposed taxonomy is regular and designed to highlight key similarities and differences in different performance methodologies. Such an insight is not available in existing literature.

## 2. BACKGROUND

Literature pertaining to performance related taxonomies has focused on specific aspects of performance evaluation-primarily on workloads and simulation methods or application specific performance issues, for example, taxonomy for imaging performance [1]. A taxonomy of hardware supported measurement approaches and instrumentation for multi-processor performance is considered in [2]. A taxonomy for test workload generation is considered in [3] that covers aspects of valid test workload generation and [4] that considers process execution characteristics. A proposal for software performance taxonomy was discussed in [5]. Work on performance simulation methods, their characteristics, and application is described in [6-9]. Another example describes specific aspects of validating pre-silicon performance verification of HubChip chipset [10]. Other related work focuses on performance verification techniques for processors and SOCs and describe specific methods used and experience from using them [11-14]. The literature, while addressing specific aspects of performance verification, addresses only part of the issues needed for complete performance assurance of a complex high performance architecture. Significant more effort is needed in producing high performance architectures and the goal of the paper is to provide a complete picture of this effort in the form of a unified taxonomy that can't be gathered through glimpses of pieces described in the literature. This paper covers key aspects of product life-cycle performance assurance methods and proposes a taxonomy to encapsulate these methods in a high level framework. We show in a later section how the proposed taxonomy covers subsets of the performance verification methods described in the literature and its application to real world high performance architectures.

Section 3 provides motivation for development of the taxonomy. Section 4 describes the proposed taxonomy. Section 5 discusses examples of application of the proposed taxonomy. Section 6 concludes the paper.

## 3. MOTIVATION FOR THE PROPOSED TAXONOMY

Product performance assurance is not a new problem and manufacturers of high performance architectures have provided snapshots of subset of the work done [10-14]. This paper unifies key methods employed in performance assurance from inception to the delivery of silicon. Such a taxonomy is useful in the following ways:

    a. It depicts how high confidence performance assurance is conducted for modern microprocessors/chipsets based on experience over several generations of products.

    b. It provides new insight into the total solution space of performance assurance methods employed for real high performance chips and a common framework within which new methods can be devised and understood.

    c. It provides a rational basis for comparison of different methods employed and shows similarities and differences between methods employed at each stage of performance assurance.

d. Exposes the complexity, flexibility, and trade-offs involved in the total task and provides a basis for identifying adequacy of performance assurance coverage obtained with a different solutions and any potential gaps that might exist that can be filled to improve coverage

e. Provides a framework for assessing risk with respect to product performance with reference to initial expectations set through planning or competitive assessment and provides a high level framework for creating a detailed performance assurance execution plan

Why is it important to look at a detailed framework for components of performance assurance? To understand this, it is useful to go through the process of specification of performance requirements and their evaluation through product life cycle:

- Performance targets for a new architecture and its derived products are set via careful planning for the time frame when it is introduced to make it competitive.
- A set of high level ideas to reach performance targets are investigated via a high level model and a subset of these ideas is selected for implementation.
- A micro-architecture for implementing the selected ideas is designed and RTL (register transfer level) model is created.
- Silicon implementing the RTL model is created and tested.

Performance evaluation is necessary at each stage to meet the set targets. The tools used for performance analysis at each stage differ greatly in their capabilities, coverage, accuracy, and speed. Table 1 shows how various attributes of performance assurance at each stage compare. A high level performance model can be developed rapidly, can project performance for modest number and sizes of workloads, stimulus can be injected and observed at fine granularity but may not capture all micro-architecture details. Performance testing with an RTL model needs longer development time, runs slow, and can project performance for a small set of workloads over short durations but captures details of the micro-architecture. Performance testing with silicon can run full set of workloads, captures all details of micro-architecture and provides significant coverage of performance space, however, ability to inject stimulus and observability of results is limited. The goals of performance testing in these stages are also different. In the high level model the goal is feature definition and initial performance projections to help reach the goals and evaluate performance trade-off vs. micro-architecture changes needed from initial definition at a later stage to see if it is still acceptable. The goal of RTL performance testing is to validate that the policies/algorithms specified by high level feature definition are correctly implemented and correlated on a preselected set of tests on key metrics and that performance is regularly regressed against implementation changes. Silicon performance is what is seen by the customer of the product and its goal is to test that the initial performance targets are met and published externally, it also provides key insights for development of next architecture via measured data with any programmable features and de-features in the chip.

Considering these differences in the capabilities and goals of performance assurance at each stage, thinking of performance in a monolithic manner does not help one easily comprehend the complete space needed to deliver high performance architectures. It is important to tackle performance assurance at each stage of development process with a clear understanding of the goals, capabilities, and limitations to understand the scope and gaps in coverage that is addressed by the proposed taxonomy.

Table 1. Comparison of attributes of performance testing with different abstraction levels.

| | HL Performance | RTL Performance | Silicon Performance |
|---|---|---|---|
| Development time | Low | Modest | High |
| Workload size and length | Modest | Short | Long |
| Stimulus Injection granule | Fine | Fine | Coarse |
| Observation Granule | Fine | Fine | Coarse |
| Result speed | Modest | Slow | Fast |
| Microarchitecture Detail captured/tested (accuracy) | Low | High | High |
| Perf space coverage | Modest | Modest | High |
| Goal | High level arch partitions, pre-si feature defn, pre-si pef projection, implementation cost vs. perf tradeoff | Validate arch policies get implemented in RTL, maintain projected performance | Validate expected silicon performance from part, provide input for next generation arch, perf over competition or next process shrink |

## 4. A TAXONOMY FOR PERFORMANCE ASSURANCE

The total performance assurance space (PA) consists of a cross product of two spaces-independent performance assurance space (IPA) and correlation performance assurance space (CPA). IPA marks the space covered by independently testing each of the three abstraction levels whereas CPA marks space covered by correlating performance between combinations of abstraction levels. Examples of IPA space performance testing includes: performance comparison with a feature on vs off, performance comparison with previous generation, performance sensitivity to key micro-architecture parameters (policies, pipeline latency, buffer sizes, bus width, speeds etc), benchmark score projections, transaction flow visual defect analysis (pipeline bubbles), idle/loaded latency and peak bandwidth measurements, multi-source traffic interference impact, etc. CPA space correlates measurements done in one space to that done in other space with comparable configurations on various metrics to identify miscorrelations and gain confidence. Coverage in both spaces is needed to get high confidence in performance. We discuss each level of abstraction and propose a taxonomy consisting of the following four components.

Let us denote:

$\alpha$ as the high level performance space,
$\beta$ as the RTL performance space,
$\gamma$ as the silicon performance space,
$\theta$ as the correlation performance assurance space (CPA), of individual spaces ($\alpha$, $\beta$, $\gamma$), then the taxonomy for performance assurance space for high performance architectures (PA) denoted by $\delta$, is given as:

$$\delta \in \{ \ \alpha \ X \ \beta \ X \gamma \ X \ \theta \} \ \text{or} \ \{\text{IPA X CPA}\} \tag{1}$$

Where X denotes a Cartesian product of individual spaces. IPA is marked by $\{ \ \alpha \ X \ \beta \ X \ \gamma \ \}$.

## 4.1. High level performance assurance space α

Figure 1 depicts high level (HL) performance assurance space. Components of the space exploit symmetry in providing coverage in all spaces to generate a regular taxonomy.

$$\alpha \in \{ \text{ Analysis method } (\lambda) \text{ X Stimulus } (\phi) \text{ X Component granularity } (\mu) \text{ X Transaction}$$
$$\text{source } (\eta) \text{ X Metric } (\rho) \text{ X Configuration } (\xi) \} \tag{2}$$

Where:

*Component granularity ($\mu$)* $\in$ {Platform, full chip, cluster, combination} (3)
*Analysis method ($\lambda$)* $\in$ {Analytical model, simulation model, emulation, combination} (4)

*Stimulus ($\phi$)* $\in$ {Complete workload/benchmark, samples of execution traces, synthetic/directed

workload, combination} (5)
*Traffic source ($\eta$)* $\in$ {Single source, multiple sources } (6)

*Metric ($\rho$)* $\in$ {Benchmark score, throughput/runtime, latency /bandwidth, meeting
area/power/complexity constraints, combination} (7)
*Configuration ($\xi$)* $\in$ {Single configuration, multiple configurations} (8)

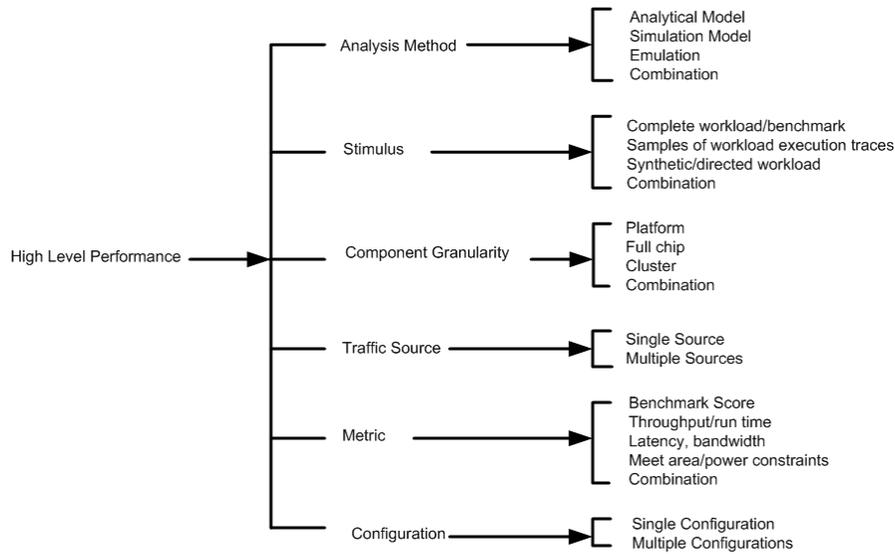

Figure 1: IPA-space of high level performance assurance

High level performance analysis may be done using analytical model, simulation model, emulation, or a combination of these methods. Analytical models are suitable for rapid high level analysis of architectural partitions when the behavior and stimulus is well understood or can be abstracted as such, simulation modeling may be trace or execution driven and can incorporate more details of the behavior to get higher confidence in performance analysis under complex behavior and irregular stimulus, emulation is suitable when an emulation platform is available and speed of execution is important. A behavioral high level simulation model may describe different units with different abstraction levels (accuracy) and gets progressively more accurate with respect to the implementation details as RTL is coded and correlated, the HL model serves as a reference in later stages. A combination method can also be used for example, a spread-sheet model that combines an analytical model with input from simulation model, if it is too expensive to simulate underlying system with adequate accuracy and speed.

We may choose to test the system at different levels of component granularity. It is possible to test at platform level (where the device under test is a component of the user platform), at full chip level where the device under test is a chip implementing the high performance architecture, for example, in a high volume manufacturing tester, at a large cluster level within the chip (for example: out of order execution unit or last level cache in the uncore), or we may target all of these depending on which pieces are critical for product performance. The test stimulus and test environment for each component granularity may differ and needs infrastructure support to create comparable stimulus, configuration etc. for performance correlation.

Stimulus may be provided in several forms depending on the device under test. We may use a complete workload execution on a high level model, short trace samples from execution of a workload (e. g. running on a previous generation platform or new arch simulator) driving a simulation model, use synthetic/directed tests to exercise a specific performance feature or a cluster level latency and bandwidth characterization. Synthetic stimulus may target for example, idle or loaded latencies, cache hit/miss and memory page hit/miss bandwidth, peak read/write interconnect bandwidth (BW) etc. Synthetic stimulus can also be directed toward testing performance of new high risk features that may span across the micro-architecture. Synthetic stimulus is targeted toward testing a specific behavior and/or metric whereas a real workload trace captures combinations of micro-architecture conditions and flows that a synthetic behavior may not generate and both are important from getting good coverage. Synthetic and real workload stimuli may converge if the workload is a synthetic kernel and traces from its execution are used in driving a simulator, however, in most cases the differentiation can be maintained. Stimulus may also be a combination of these stimuli. The selected method depends on speed of execution of the model, and the importance of the metric and workloads.

For traffic sources, depending on the device under test, we may test with a single traffic source or a combination of traffic sources. Examples of a single traffic sources are CPU multi-core traffic, integrated graphics traffic, or IO traffic that might be used to characterize core, graphics, IO performance with a new feature. We may have a combination of above traffic sources to find interesting micro-architecture performance bottlenecks. Examples of such bottlenecks include for example, buffer sizes, forward progress mechanisms, coherency conflict resolution mechanisms.

Various metrics are used in evaluation. If the benchmark can be run on the HL model/silicon, a benchmark score is used. If components of the benchmark or short traces of workload execution are used, throughput (CPI) or run time is used. If performance testing is targeted to a specific cluster, we may use latency of access or bandwidth to the unit as a metric.  For a performance feature to be viable, it needs to also meet area, power, and complexity constraints in implementation. An addition of a new feature may need certain die area and incur leakage and dynamic power that impacts TDP (thermal design point) power and battery life. Based on the performance gain from a new feature and impact on the area/power, a feature may or may not be viable depending on the product level guidelines and needs to be evaluated during HL and RTL performance stages. Design/validation complexity of implementing the performance feature is a key constraint for timely delivery. We may use a combination of these metrics depending on the evaluation plan.

There may be more than one product configuration supported with a given architecture. Several possibilities exist: do complete performance testing on all configurations, a subset of the performance testing on all configurations, or a subset of the performance testing on a subset of configurations that differ in key ways to trade off effort against performance risk. The exact configurations and the performance testing with each configuration depends on the context, the proposed taxonomy differentiates between how much testing is done for each. An example of multiple configurations for a core/uncore is use in several desktop, mobile, server configurations that differ in key attributes (cache size, number of cores, core/uncore frequency, DRAM speed/size/channels, PCI lanes etc.).

Not all combinations generated in the HL space are either valid, feasible, or equally important. For example, although in principle one could specify an analytical model at platform granularity to measure benchmark score, creating such a model with desired accuracy may not be feasible. Performance testing with one configuration and traffic source may be more extensive than other combinations due to the significance attached to those tests. A performance architect will specify relevant components of the space that are deemed significant in a performance assurance execution plan. We do not enumerate key combinations as their significance differs depending on the context.

## 4.2. RTL  performance assurance space $\beta$

Figure 2 depicts the RTL performance assurance space.

$\beta \in \{$ Stimulus $(\phi)$ X Component granularity $(\mu)$ X transaction source $(\eta)$ X Metric $(\rho)$ X configuration $(\xi)$ $\}$       (9)

Where:

*Stimulus ($\phi$)* $\in \{$ Samples of execution traces, synthetic/directed workload, combination$\}$ (10)

*Component granularity ($\mu$)* $\in \{$ Full chip, cluster, combination$\}$       (11)

*Traffic source ($\eta$)* $\in \{$Single source, multiple sources $\}$       (12)

*Metric ($\rho$)* $\in \{$Throughput/runtime, latency /bandwidth, meeting area/power/complexity constraints, combination$\}$       (13)

*Configuration ($\xi$)* $\in \{$Single configuration, multiple configurations$\}$       (14)

Components of RTL performance assurance space are symmetric with the high level components except for the following key differences arising from differences in environments. Performance testing is done on RTL model that generally runs slow since it captures micro-architecture details. Running large benchmarks is thus generally hard without a large compute capacity and it is best to use short workload test snippets or directed tests. The execution results may be visually inspected or measured using performance checker rules on result log files.

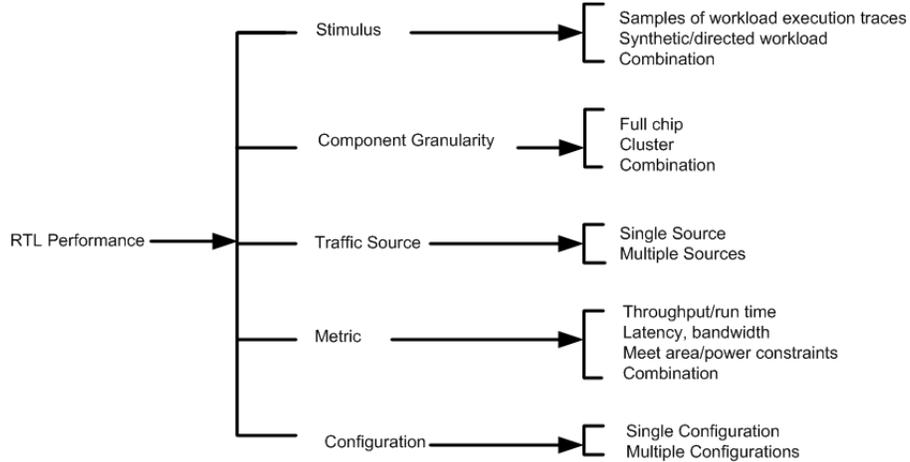

Figure 2: IPA-space of RTL performance assurance

## 4.3. Silicon performance assurance space γ

Figure 3 depicts silicon performance assurance space.

γ ∈ { Stimulus (φ) X Component granularity (μ) X transaction source (η) X Metric (ρ) X configuration (ξ) } (15)

Where:

*Stimulus (φ)* ∈ {Complete workload/benchmark, synthetic/directed workload, combination} (16)

*Component granularity (μ)* ∈ {Platform, full chip, combination} (17)

*Traffic source (η)* ∈ {Single source, multiple sources } (18)

*Metric (ρ)* ∈ {Benchmark score , Throughput/runtime, latency /bandwidth, combination} (19)

*Configuration (ξ)* ∈ {Single configuration, multiple configurations} (20)

Components of silicon performance are symmetric to other spaces with notable differences related to accessibility/observability notes earlier. Thus for devices under test, stimulus component granularity is limited to full chip/platform.

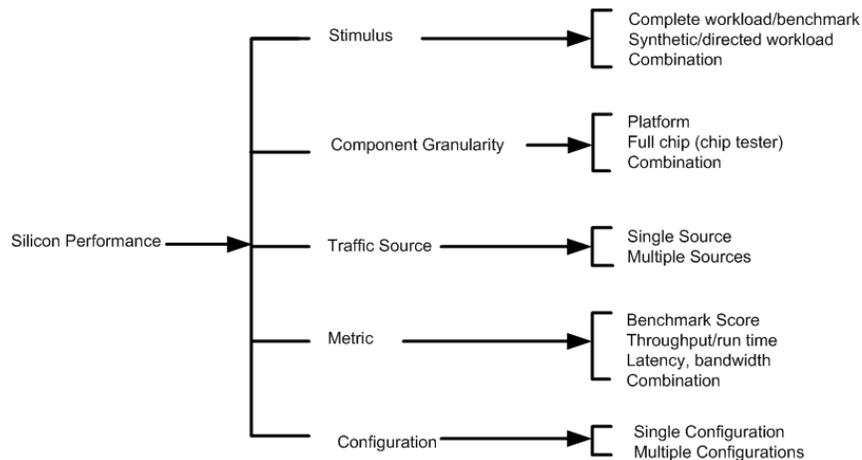

Figure 3: IPA-space of silicon performance assurance

## 4.4. Correlation performance Assurance (CPA) Space θ

Figure 4 shows four components of CPA using definitions symmetric to IPA space:
Let τ denote the correlation space between RTL and High level performance
Let ϖ denote the correlation space between High level and silicon performance
Let ∂ denote the correlation space between RTL and silicon performance
Let Ω denote the correlation space between HL, RTL, and silicon performance
Then CPA θ is given as:

$$\theta \in \{ \tau \text{ X } \varpi \text{ X } \partial \text{ X } \Omega \ \} \tag{21}$$

$$\tau , \varpi , \partial , \Omega \ \in \{ \text{ Stimulus } (\phi) \text{ X Component granularity } (\mu) \text{ X transaction source } (\eta) \text{ X Metric}$$

$$(\rho) \text{ X configuration } (\xi) \ \} \tag{22}$$

**For τ:**
*Stimulus (φ)* ∈ {Samples of execution traces, synthetic/directed workload, combination}
*Component granularity (μ)* ∈ {Full chip, cluster, combination}
*Traffic source (η)* ∈ {Single source, multiple sources }
*Metric (ρ)* ∈ {Throughput/runtime, latency /bandwidth, area/power/complexity constraint, combination}
*Configuration (ξ)* ∈ {Single configuration, multiple configurations}

**For ϖ:**
*Stimulus (φ)* ∈ {Complete workload/benchmark, synthetic/directed workload, combination}
*Component granularity (μ)* ∈ {Platform, full chip, combination}
*Traffic source (η)* ∈ {Single source, multiple sources }
*Metric (ρ)* ∈ { Benchmark score , Throughput/runtime, latency /bandwidth, combination}
*Configuration (ξ)* ∈ {Single configuration, multiple configurations}

**For ∂:**
*Stimulus (φ)* ∈ { Synthetic/directed workload}
*Component granularity (μ)* ∈ {Full chip}
*Traffic source (η)* ∈ {Single source, multiple sources }
*Metric (ρ)* ∈ {Throughput/runtime, latency /bandwidth, combination}
*Configuration (ξ)* ∈ {Single configuration, multiple configurations}

**For Ω:**
*Stimulus (φ)* ∈ {Synthetic/directed workload}
*Component granularity (μ)* ∈ {Full chip}
*Traffic source (η)* ∈ {Single source, multiple sources }
*Metric (ρ)* ∈ {Throughput/runtime, latency /bandwidth, combination}
*Configuration (ξ)* ∈ {Single configuration, multiple configurations}

CPA space denotes the part of the total coverage that is obtained by correlating between IPA spaces using comparable stimulus, metric, traffic sources, components, and configurations. This coverage is necessary because we are not able to test everything in individual spaces due to limitations discussed earlier and correlation space improves that coverage. In CPA space, high priority is on correlating the performance of the RTL model with the high level model. The high level model runs fast enough and can be used to project benchmark level performance and if the two models correlate, the high level model serves as a good proxy for what we may expect for RTL benchmark level projection. The significance of each correlation space may differ. We have discussed individual components of each space earlier and their definition is not repeated here for brevity.

The PA taxonomy for high performance architectures provides a new way to look at the complete performance assurance space that is easily understood and extended using a well defined and regular set of criteria. The criteria used in defining performance assurance space are represented by a key set of issues that an architect would need to resolve while designing the solution. This does not mean it includes every possible issue as a taxonomy based on such an endeavor would be unwieldy. The selected criteria are relevant to all abstraction levels, capture key issues that need to be addressed, and any significant differences between the levels can be isolated using the criteria. We discuss application of this taxonomy in the next section.

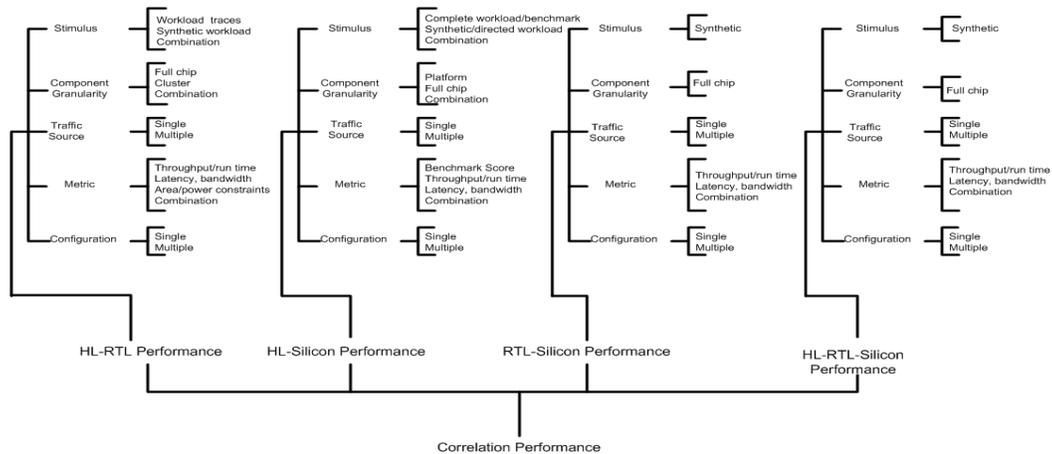

Figure 4: Correlation performance assurance space (CPA)

## 5. APPLICATION AND CONSIDERATIONS

### 5.1. Solution Spaces and Coverage

Figure 5. shows that the proposed taxonomy partitions the total performance assurance space into seven distinct spaces. The IPA is marked by spaces 1, 2, and 3. CPA space is marked by spaces 4, 5, 6, 7 that overlap IPA spaces. Table 2 illustrates high level characteristics of each space and shows what areas they may cover. The table is meant to be illustrative and not an exhaustive coverage of each space. For example, if synthetic/directed stimulus is missing from the selected solution in all components and instead have only real workload/traces for stimulus, there may be a hole in testing peak bandwidth of key micro-architecture components. If synthetic/directed tests were present only in silicon performance, then the testing gap may propagate until silicon through HL and RTL performance and may be expensive to fix later. Similar consideration applies to dropping testing of a high risk feature from one or more of the spaces using synthetic/directed tests. In these cases, real workload traces may not find a performance problem with the feature without explicit directed testing and may result in a potential performance coverage hole. Similar coverage comments apply to CPA space in the table depending on what coverage is sought. For detailed gap/risk assessment, more details of each component of the solution need to be specified in an assurance plan and the combinations reviewed over the PA space, for example, models needed for evaluation, list of workloads, details of synthetic tests targeting specific behaviors/features, details of clusters, traffic sources, detailed metrics and configurations.

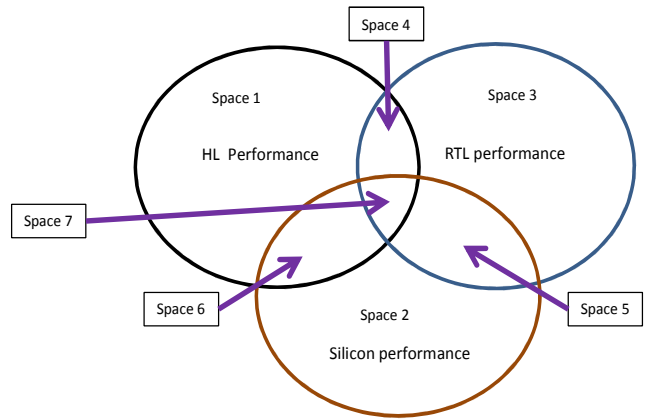

Figure 5: Solution Spaces of Performance Assurance Methods

Depending on a product's life stage and goals, coverage in all spaces may not be equally important. For example, for a product design to deliver expected performance covering space 4 (RTL performance correlation with high level model) may be more important than covering space 5 that would test micro-architecture defeatures, hardware performance counters/events etc. Similarly, space 7 may be higher priority than space 5 and one could make coverage, effort tradeoffs/prioritization that way.

Table 2: Example of coverage provided by each solution space

|  | **Performance Validation Space Coverage** |
|---|---|
| Space 1 | (IPA-☐) Rapid HL Performance analysis of architecture partitions, new features, less micro-architecture details (more refined as micro-architecture is defined), set product performance projections/expectations |
| Space 2 | (IPA-☐)Silicon performance. Product performance projections published for various benchmarks with silicon implementation or measuring and comparing performance with competitive products, tune parameters to optimize performance (BIOS setting) |
| Space 3 | (IPA-☐) Testing of RTL performance with short tests to get confidence in performance after functional coding at unit/cluster level with  details of micro-architecture implemented, transaction flow inspection for defects (bubbles) |
| Space 4 | (CPA-τ) Verify that RTL implemented algorithms specified in the architecture specification derived from high level analysis by correlating with HL model with short tests and snippets of workloads. Validate and correlate  changes in micro-architecture required from implementation complexity and their performance impact |
| Space 5 | (CPA-∂)Test/validate cases that have performance impact and needs details of micro-architecture not implemented in HL model, examples- product defeatures, rare architecture/micro-architecture corner cases with short full chip tests, hardware performance counters and events, other performance observability hooks |
| Space 6 | (CPA-ϖ)Test full benchmark execution and correlate silicon performance to that projected with a high level model to see if it meets targets when the full implementation is considered, provides a method for correlation of pre and post silicon measurements and validation of pre silicon methodologies,  also useful for providing input for next generation CPUs with targeted studies of features and defeatures |
| Space 7 | (CPA-Ω)This is intersection of all three methods and used to test performance pillars in all cases. For example running full chip micros/directed tests for key component latencies and bandwidths  and high risk features which can be regularly tracked in a regression suite as the RTL and silicon steppings change |

We illustrate below application of the taxonomy to performance verification described in the literature and then show more complete examples of application of taxonomy to specific examples of high performance Intel processors and MCH (memory controller hub) chipsets. These examples depict how the performance verification work in the literature can be described under the proposed framework and how the taxonomy extends to testing with real chips.

## 5.2. Application Examples

Application of the taxonomy to work done in the literature is shown only to the specific methods discussed in these papers and does not reflect on whether the products described were limited to testing shown here. We consider example discussed in [10, 11, 12-13]. In 10, *Doering et al* consider performance verification for high performance PERCS Hub chip developed by IBM that binds several cores and IO. This work largely relates to high level (analytical(queue)+simulation(OMNET) ) and VHDL RTL correlation for the chipset. In the proposed taxonomy, the work described in the paper would be classified under CPA space and HL-RTL correlation (τ) branch of CPA as follows:

*HL-RTL Correlation ∈ {Stimulus=Trace driven, Component granularity=full chip, Traffic source=multiple, Metric=multiple (latency, throughput), Configuration= single}*

In 11, *Holt et al* describe system level performance verification of multi-core SOC. Two methods of performance verification have been described in this paper-*top down* and *bottom up* verification. Under the proposed taxonomy, the top down performance verification would be described under IPA HL performance assurance (□ branch whereas bottom up performance verification would be described under CPA HL-RTL correlation (τ) branch as follows:

*(**Top down**) IPA HL performance ∈ {Analysis method= emulation, Stimulus=synthetic, Component granularity=full chip, Traffic source=multiple, Metric=combination (latency/BW, throughput), Configuration= multiple}*

*(**Bottom up**) CPA HL-RTL correlation ∈ {Stimulus=synthetic, Component granularity=full chip, Traffic source=multiple, Metric=combination (unloaded latency, throughput), Configuration= single}*

In 12, 13 *Bose et al* have described architecture performance verification of IBM's PowerPC ™ Processors. Under the proposed taxonomy, the work described here would be included in the CPA space and HL-RTL correlation (τ) branch of CPA as follows:

*HL-RTL Correlation ∈ {Stimulus=combination, Component granularity=full chip, Traffic source=single (CPU core), Metric=combination (latency/BW, throughput), Configuration= Multiple (Power3, Power4)}*

These examples show that the performance verification work in the literature focuses on subset of the PA space and there is no clear definition of the whole space. The proposed taxonomy achieves two goals-describes the total space and provides a consistent terminology to describe parts of the total space. The classification above also shows high level similarities and differences in the methods used in these cases.

Next, we show application of the proposed taxonomy to three examples: IA™ CPU core, MCH chipset, and a memory controller cluster in Table 3. The first table shows IPA space and the second table shows CPA space. The taxonomy mapping for each example is illustrative and other solutions are possible depending on the context.

For IPA HL core performance, a combination of analytical model during early exploration and a simulation model of the architecture are used. The stimulus is a combination of directed tests for specific latency/BW characterization and real workload benchmark traces for high quality coverage. The directed tests also cover new features introduced in the architecture. The testing is done as a combination of cluster and full chip granularity. Single source of traffic is IA™ core workloads/traces and the measurement granularity is a combination of benchmark score, throughput/run time, and latency and BW of targeted units. Since the core is used in multiple configurations (desktop, mobile, server), testing is done with multiple configurations. For core RTL testing, similar considerations apply as high level testing except that the metrics are run time/throughput and latency bandwidth combination and stimulus contains a combination of traces and synthetic workload. For core Silicon testing, the stimulus consists of combination of complete workload and directed full chip tests, and component granularity is platform and full chip. Other considerations for metric and configurations with RTL and silicon performance are comparable to that of HL performance.

For IPA MCH chipset performance testing (in chipset column), one significant difference is in the traffic source. The core had a single source of traffic, MCH binds multiple sources that includes cores, IO, graphics. The performance testing for MCH  is done with multiple sources of transactions and combination of metrics. If the MCH functionality is integrated into an uncore or a SOC, it would have a comparable IPA scheme.

A memory controller (MC) is a cluster within the uncore or MCH and its performance testing is shown as the third example. It can also be left as a part of uncore cluster/MCH testing if considered adequate. In this example, we consider memory controller as a modular component that may be used for targeting more than one architecture and thus needs to be independently tested for high confidence. High level IPA testing of a memory controller is done with a simulation model and synthetic micros directed at performance aspects of a memory controller that test core timings, turnarounds, latency, and BW under various read write mixes and page hit/miss proportions. It can be tested with multiple traffic sources with different memory configurations (number of ranks, DIMMS, speeds, timings etc.). For silicon testing, memory controller performance is tested as a combination of synthetic workloads and benchmarks (streams) etc.

CPA space for all four components is shown in the second table. For example, for a CPU core HL-RTL correlation, a combination of short real workload traces along with synthetic workloads is tested on the HL model and RTL at full chip and cluster combination. The workload source is an IA core and a combination of metrics throughput (for workload traces) and latency/BW (with synthetic workload) is used for correlation. This correlation is done on multiple configurations. For HL-silicon correlation, combination of full chip latency/BW micros and benchmarks are run and correlation is done for benchmark scores and latency/BW metrics. This correlation also helps improve the HL model accuracy and a useful reference for development of next generation processors. For CPU core, RTL silicon correlation is done on a single configuration whereas other three correlations are done on multiple configurations. This illustrates an example of trading effort vs. coverage at a low risk since RTL silicon correlation covers uncommon cases from performance perspective and get adequate testing on a single configuration. The HL-RTL-Silicon correlation testing is done with targeted synthetic full chip micros that test the core metrics that are key for product performance and the testing is done at full chip with combination of throughput and latency/BW metrics in multiple configurations. Similar considerations apply to chipset and memory controller CPA space.

Table 3: Example of application of taxonomy to real world examples

| IPA | | CPU core | Chipset | Memory Controller unit |
|---|---|---|---|---|
| **High Level Testing** | Analysis method | Combination | Simulation | Simulation |
| | Stimulus | Combination | Combination | Synthetic |
| | Component Granularity | Combination | Combination | Cluster |
| | Traffic src | Single | Multiple (IA/IO/GFX) | Multiple |
| | Metric | Combination | Combination | Latency/BW |
| | Configs | Multiple | Multiple | Multiple |
| **RTL** | Stimulus | Combination | Combination | Synthetic |
| | Component Granularity | Combination | Combination | Cluster |
| | Traffic src | Single | Multiple | Multiple |
| | Metric | Combination | Combination | Latency/BW |
| | Configs | Multiple | Multiple | Multiple |
| **Silicon** | Stimulus | Combination | Combination | Combination |
| | Component Granularity | Combination | Combination | Platform |
| | Traffic src | Single | Multiple | Multiple |
| | Metric | Combination | Multiple | Latency/BW |
| | Configs | Multiple | Multiple | Multiple |

| CPA | | CPU core | Chipset | Memory Controller unit |
|---|---|---|---|---|
| **HL-RTL** | Stimulus | Combination | Synthetic | Synthetic |
| | Component Granularity | Combination | Full chip | Cluster |
| | Traffic src | Single | Multiple (IA/IO/GFX) | Multiple |
| | Metric | Combination | Combination | Latency/BW |
| | Configs | Multiple | Single | Multiple |
| **HL-Silicon** | Stimulus | Combination | Synthetic | Synthetic |
| | Component Granularity | Combination | Full chip | Full chip |
| | Traffic src | Single | Multiple | Multiple |
| | Metric | Combination | Combination | Latency/BW |
| | Configs | Multiple | Single | Multiple |
| **RTL-Silicon** | Stimulus | Synthetic | Synthetic | Synthetic |
| | Component Granularity | Full chip | Full chip | Full chip |
| | Traffic src | Single | Single | Single |
| | Metric | Combination | Throughput/run time | Latency/BW |
| | Configs | Single | Single | Single |
| **HL-RTL-Silicon** | Stimulus | Synthetic | Synthetic | Synthetic |
| | Component Granularity | Full chip | Full chip | Full chip |
| | Traffic src | Single | Multiple | Single |
| | Metric | Combination | Latency/BW | Latency/BW |
| | Configs | Multiple | Single | Multiple |

# 6. CONCLUSIONS

This paper presented a systematic approach to the complex problem of performance assurance of high performance architectures manufactured in high volume based on methods successfully deployed over several generations of Intel cores/chipsets in a unified taxonomy. The taxonomy extensively considers performance assurance through three key stages of a product that include high level product performance, RTL performance, and silicon performance and has not been discussed in the literature previously. The proposed taxonomy incorporated capabilities and limitations of performance tools used at each stage and helps one construct a complete high level picture of performance testing that needs to be done at each stage. An application of the taxonomy to examples in the literature and real world examples of a CPU core, MCH chipset, and memory controller cluster are shown.

The key advantages of proposed taxonomy are: it shows at high level where the performance assurance methods need to be different, it makes one think through all phases of a product starting from high level until silicon, enumeration of the taxonomy in a detailed performance assurance execution plan identifies if there are holes in the performance testing that either need to be filled or concomitant risk is appropriately assessed. The taxonomy helps with resource planning and mapping and delivering a successful high performance product.

The proposed taxonomy has been successfully used in performance assurance of Intel's Nehalem/Westmere CPUs and several generations of chipsets. This systematic approach has been instrumental in identifying many pre-silicon performance issues early on and any corner cases identified in silicon due to several cross checks embedded in the methodology. It has helped creating a rigorous performance assurance plan. The proposed work is new and should be of interest to manufacturers of high performance architectures.

# 7. REFERENCES

[1] Don Williams, Peter D. Burns, Larry Scarff, (2009) "Imaging performance taxonomy"; Proc. SPIE 7242, 724208; doi:10.1117/12.806236, Monday 19 January 2009, San Jose, CA, USA

[2] Mink, A.; Carpenter, R.J.; Nacht, G.G.; Roberts, J.W.; (1990) "Multiprocessor performance-measurement instrumentation", Computer, Volume: 23 , Issue: 9, Digital Object Identifier: 10.1109/2.58219, Page(s): 63 - 75

[3] Mamrak, S.A.; Abrams, M.D, (1979) "Special Feature: A Taxonomy for Valid Test Workload Generation "; Computer, Volume: 12, Issue: 12, Digital Object Identifier 10.1109/MC.1979.1658577, Page(s): 60 – 65

[4] Oliver, R.L.; Teller, P.J.; (1999) "Are all scientific workloads equal?", Performance, Computing and Communications Conference, 1999. IPCCC '99. IEEE International, Digital Object Identifier: 10.1109/PCCC.1999.749450, Page(s): 284 - 290

[5] Mary Hesselgrave , (2002) "Panel: constructing a performance taxonomy", July 2002   WOSP '02: Proceedings of the 3rd international workshop on Software and performance

[6] S. Mukherjee, S Adve, T. Austin, J. Emer, P. Magnisson, (2002) "Performance simulation tools", Computer , Issue Date : Feb 2002, Volume : 35 , Issue:2, On page(s): 38, Digital Sponsored by : IEEE Computer Society

[7] S. Mukherjee, S. Reinhardt, B. Falsafi, M. Litzkow, S. Huss-Lederman, M. Hill, J. Larus, and D. Wood, (2000) "Fast and portable parallel architecture simulators: Wisconsin wind tunnel II", IEEE Concurrency, vol. 8, no. 4, pp. 12–20, Oct.–Dec. 2000.

[8] Heekyung Kim , Dukyoung Yun , (2009) "Scalable and re-targetable simulation techniques for systems", Proceeding CODES+ISSS '09 Proceedings of the 7th IEEE/ACM international conference on Hardware/software codesign and system synthesis , NY 2009

[9] Hoe, James C.; Burger, Doug; Emer, Joel; Chiou, Derek; Sendag, Resit; Yi, Joshua; (2010) "The Future of Architectural Simulation", Micro, IEEE, Volume: 30 , Issue: 3, Digital Object Identifier: 10.1109/MM.2010.56, Page(s): 8 - 18

[10] Andreas Doering and Hanspeter Ineichen, "Visualization of Simulation Results for the PERCS HubChip Performance Verification", Proc. SIMUTooLs 2011, 4th ICST conf on simulation tools and techniques, March 21-25, Barcelona, Spain

[11] Holt, J.; Dastidar, J.; Lindberg, D.; Pape, J.; Peng Yang; "System-level Performance Verification of Multicore Systems-on-Chip", Microprocessor Test and Verification (MTV), 2009 10th International Workshop on Digital Object Identifier: 10.1109/MTV.2009.10, Publication Year: 2009 , Page(s): 83 - 87

[12] Surya, S.; Bose, P.; Abraham, J.A.; "Architectural performance verification: PowerPC ", Computer Design: VLSI in Computers and Processors, 1994. ICCD '94. Proceedings., IEEE International Conference on Digital Object Identifier: 10.1109/ICCD.1994.331922 , Publication Year: 1994 , Page(s): 344 - 347

[13] Bose, P.; " Ensuring dependable processor performance: an experience report on pre-silicon performance validation ", Dependable Systems and Networks, 2001. DSN 2001. International Conference on, Digital Object Identifier: 10.1109/DSN.2001.941432, Publication Year: 2001 , Page(s): 481 - 486

[14] Richter, K.; Jersak, M.; Ernst, R.; "A formal approach to MpSoC performance verification", Computer, Volume: 36 , Issue: 4, Digital Object Identifier: 10.1109/MC.2003.1193230, Publication Year: 2003 , Page(s): 60 – 67.

## Authors

Hemant Rotithor: Received his M.S and Ph.D. in Electrical and computer Engineering from IIT Bombay and University of Kentucky. He taught at Worcester Polytechnic Institute, worked at DEC on compiler performance analysis; he is currently working at Intel Corporation in Hillsboro Oregon  in the microprocessor architecture group. At Intel Corporation, he has worked on performance of many generations of microprocessors and chipsets. He has several patents issued in the area of uncore microarchitecture performance, memory scheduling, and power management. He has published papers on performance analysis, and distributed computing, and validation.

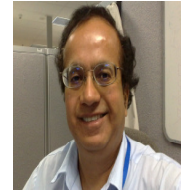